# How collective phenomena impact CO$_2$ reactivity and speciation in different media

*Daniela Polino[a, b], Emanuele Grifoni[a, b], Roger Rousseau[c], Michele Parrinello[a, b, d] and Vassiliki-Alexandra Glezakou\*[c]*

[a] Department of Chemistry and Applied Biosciences, ETH Zurich, c/o USI Campus Via Giuseppe Buffi 13, CH-6900 Lugano Switzerland.

[b] Facoltà di Informatica, Istituto di Scienze Computazionali, Università della Svizzera Italiana, Via Giuseppe Buffi 13, CH-6900 Lugano Switzerland.

[c] Pacific Northwest National Laboratory, 902 Battelle Blvd, PO Box 999, MSIN K1-83, Richland WA 99352.

[d] Istituto Italiano di Tecnologia, Via Morego 30, 16163 Genova, Italy.

**ABSTRACT:** $CO_2$ has attracted considerable attention in the recent years due to its role in the greenhouse effect and environmental management. While its reaction with water has been studied extensively, the same cannot be said for reactivity in supercritical $CO_2$ phase, where the conjugate acid/base equilibria proceed through different mechanisms and activation barriers. In spite of the apparent simplicity of the $CO_2 + H_2O$ reaction, the collective effect of different environments has drastic influence on the free energy profile. Enhanced sampling techniques and well-tailored collective variables provide a detailed picture of the enthalpic and entropic drivers underscoring the differences in the formation mechanism of carbonic acid in the gas, aqueous and supercritical $CO_2$ phases.

# 1. INTRODUCTION

In recent years, a number of $CO_2$ management technologies has been developed to mitigate the adverse effects of anthropogenic emissions. Promising solutions involve carbon capture and sequestration (CCS) practices,[1] where geological formations are exposed to captured $CO_2$ that is chemically and permanently fixated. In the past decade, a number of solvent systems has been developed for selective removal of $CO_2$[2] from point sources in the form of supercritical $CO_2$ (sc$CO_2$). Supercritical fluids possess intriguing properties due to their short- and long-range inhomogeneities that researchers believe are at the center of their unusual reactivity.[3] In contrast to water mediated reactions,[4] reactivity in sc$CO_2$ has not received much attention. The well-defined concepts of solubility or activity in water do not necessarily have corresponding thermodynamic meaning, and the widely accepted reaction between $H_2O$ and $CO_2$ in aqueous environments, eq. (1), may not be proceeding the same way in sc$CO_2$, eq. (2), due to the local inhomogeneities of this condensed phase:

$$CO_2(aq) + H_2O(aq) \rightleftharpoons H_2CO_3(aq) \rightleftharpoons HCO_3^-(aq) \rightleftharpoons CO_3^{2-} \qquad (1)$$

$$scCO_2 + H_2O(sc) \rightleftharpoons H_2CO_3(sc) \qquad (2)$$

Theory and simulation are uniquely equipped to provide molecular level understanding of the structural subtleties due to both local and long range fluctuations. The microsolvation of $CO_2$ with $H_2O$ has been studied quite extensively by means of quantum mechanical methods such as density functional theory (DFT), Hartree-Fock (HF) or post-Hartree-Fock methods (e.g., perturbation or coupled cluster theories). Jena and Mishra[5] studied the $CO_2$-$H_2O$ formation and rotational barriers in water clusters, concluding that $CO_2$ solvation is likely to include more than 8 $H_2O$ molecules, and that use of higher correlation methods and basis sets alone does not improve the agreement with experiment. Nguyen et al.[6] also studied the microsolvation of $CO_2$ by $nH_2O$ (n=1-4) with MP2 and CCSD(T) methods, and found that the barrier to $H_2CO_3$ formation with 1 $H_2O$ is as high as 50 kcal/mol. Addition of 4 $H_2O$ molecules successively lowers the energy barrier to ~20–27 kcal/mol and the free energy barrier to ~19–23 kcal/mol depending on the local geometries of the waters, where three of them directly participate in the transition state. Loerting et al.[7,8] reported similar values, 52.6 kcal/mol and 43.55 kcal/mol for $H_2CO_3$ dissociation, while adding one $H_2O$ reduced the barriers to 34.7 and 27.13 kcal/mol, and adding two $H_2O$ to 31.6 and 24.01 kcal/mol, respectively. Tautermann et al.[9] studied the stability and decomposition rate of $H_2CO_3$ and determined barriers of 44.8 kcal/mol and 52.8 kcal/mol for the $H_2CO_3$ formation. As other studies have concluded, addition of more waters helps reduce both the forward and reverse barriers, where the additional waters assist in a Grotthus-like proton transport.

In all studies, it is obvious that the dynamic role of actively partaking and spectator water molecules greatly influences both the forward ($H_2CO_3$ formation) and reverse ($H_2CO_3$ decomposition) reactions. Kumar et al.[10] used metadynamics to evaluate the energetics, conformational changes and gas phase dissociation of carbonic acid yielding a free energy barrier of 37.1 kcal/mol. Finally, Gallet et al.[11] used metadynamics to study the interactions of $CO_2$ with water molecules and the dynamic role of participating waters in the formation/decomposition of carbonic acid. These studies, albeit first of their kind, provided only a qualitative sampling of the free energy surface due to the relatively low number of reactive events. In fact, Kumar et al. observed only one reactive event per simulation, while Gallet et al. performed a longer simulation that allowed them to observe about 5 re-crossing events (10 reactive events) in 200 ps of simulation time.

The present work was motivated by recent studies of water-bearing supercritical $CO_2$. In 2007, Saharay and Balasubramanian first used ab initio molecular dynamics to study the structure and electrostatics of a solitary water molecule in supercritical $CO_2$ at three different densities.[12] Their study shows that the instantaneous $CO_2$-$H_2O$ interactions give rise to the formation of H-bonds, an increase in the water dipole moment, and larger $CO_2$ deviations from linearity. Sc$CO_2$ has a characteristic structure where each $CO_2$ is surrounded by ~6 other $CO_2$ molecules around the carbon in a distorted T-shaped configuration and 3+3 $CO_2$ molecules coordinating the oxygen poles. In 2009, Saharay and Balasubramanian performed a study of one $H_2O$ in sc$CO_2$ and reported on the structure, electronic and dynamic properties of $H_2O$ in this condensed phase, but did not examine any reactivity.[13] In 2010 Glezakou et al. examined the structure, dynamics, and vibrational spectra of sc$CO_2$/$(H_2O)_n$, where n=0-4 at the carbon sequestration relevant conditions (density of 0.74 g/cm$^3$ and T=318.15 K) using ab initio molecular dynamics.[14] The analysis showed that the waters do not disrupt the sc$CO_2$ structure and that the strongest interactions between $CO_2$/$H_2O$ occur in the case of monomeric water, where the water is mostly found in between the equatorial and polar coordinating $CO_2$ molecules, forming dynamic hydrogen bonds with the oxygens. These findings led us to believe that the interactions and chemistry of $CO_2$/$H_2O$ in sc$CO_2$ could be quite distinct and different than in aqueous environments.

This work examines and compares the $CO_2$/$H_2O$ reactivity in gas, sc$CO_2$ and aqueous phases by means of metadynamics to sample the collective fluctuations that lead to a reaction between $CO_2$ and $H_2O$ and form carbonic acid and its derivative ions. As we will show, the choice of the appropriate collective variables that can describe both the formation and decomposition of carbonic acid and re-crossing events, is a non-trivial process. In the case of

scCO$_2$, it has to properly describe the oxygen scrambling[15] and in the case of water, both the oxygen scrambling and proton hopping.[16] At the same time, this method allows us to compute the free energy landscape for this reaction, and assess the role and source of entropy for the same reaction in the different environments. To our knowledge, this is the first comprehensive study of this fundamental reaction that compares the reactivity in these distinct environments relevant to common environmental processes, such as CO$_2$ sequestration, CO$_2$ conversion, and corrosion in scCO$_2$ environments.

## 2. COMPUTATIONAL MODELS AND METHODS

**2.1. Molecular dynamics.** Ab initio molecular dynamics simulations were carried out using the CP2K package[17] driven by the PLUMED2 code.[18,19] The Born-Oppenheimer forces were used to propagate the dynamics of nuclei, with a convergence criterion of $5 \times 10^{-6}$ a.u. for the optimization of the wavefunction. In this scheme, only valence electrons were explicitly considered. More specifically, the Khon-Sham orbitals were represented by a Gaussian basis set, while a supplementary plane-wave basis was used to represent the electron density. We opted for the MOLOPT-DZVP (2s2p1d/2s1p Gaussian basis set)[20] and a plane wave cutoff of 400 Ry. Core electrons were treated using the Goedecker-Teter-Hutter (GTH) pseudopotentials.[21] The revPBE exchange-correlation density functional[22] augmented by Grimme's third generation dispersion corrections[23] was used to account for non-covalent effects. The time-step for the integration of the equations of motion was set to 1 fs. Simulations were carried out within the NVT ensemble at 323 K using the stochastic velocity-rescaling of Bussi et al.[24] For the three environments investigated (gas, scCO$_2$, and water phase), we used a periodic cubic box of 12 Å with 1 CO$_2$ and 1 H$_2$O, a 20.5 Å with 1 H$_2$O and 63 CO$_2$ molecules to simulate super-critical CO$_2$ (density 0.74 g/cm$^3$ and T=318.15 K) and a box of 12.57 Å with 1 CO$_2$ and 63 H$_2$O to reproduce the water environment (density 0.998 g/cm$^3$). Two additional simulations were also performed starting with the formed product ($H_2CO_3$) in both scCO$_2$ and water boxes to explore the phase space of the product. All systems were initially equilibrated for ~20 ps (gas phase system) and ~100 ps for the condensed systems and analyses on the unbiased systems after discarding the first 20–30 ps. We used the equilibrated systems as starting points for the metadynamics simulations. Analysis was performed on trajectories of 1.5 ns for the gas phase system, ~600 ps for the scCO$_2$, and ~2 ns for the water system, after we observed 8–10 crossings per system, and after discarding the initial steps before the first crossing.

**2.2 Construction of collective variables.** Since the formation of carbonic acid is an activated process, we accelerated the dynamics of this event by means of metadynamics.[25-28] Metadynamics enhances the sampling of rare events by introducing a history dependent bias potential function of a set of collective variables (CVs). This repulsive potential can be written as a sum of Gaussians deposited along the system trajectory in the CV space, thus discouraging the system from revisiting already sampled configurational space.

Scheme 1 depicts the gas phase reaction between $CO_2$ and $H_2O$, along with the transition state and products to illustrate the basic features of the reactants, products and definition of the collective variables. The product, $H_2CO_3$ can exist in different conformers that are identifiable as stationary structures,[7,10,29,30] and by spectroscopic methods.[31,32]

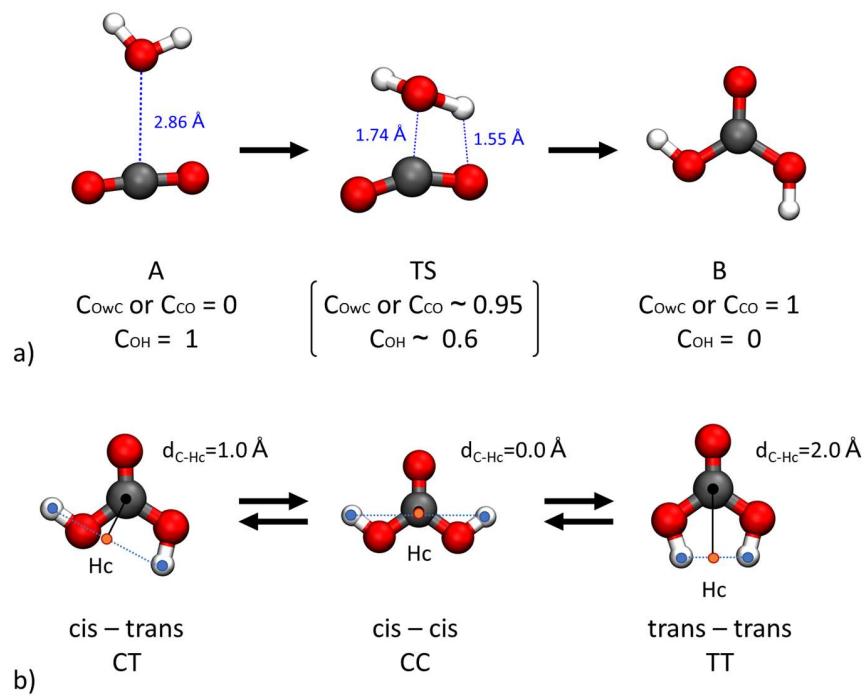

**Scheme 1.** (a) Reaction of carbonic acid formation $CO_2+H_2O \rightarrow H_2CO_3$ and (b) carbonic acid conformers, where $C_i$ defines the coordination number for the water O and H with the C and O of $CO_2$ respectively.

The definition of a set of collective variables that is able to describe the process in both directions is non-trivial. The coordination number-based path collective variables introduced by Pietrucci and Saitta[33,34] are considered prime candidates for this task. Recently, in fact, they have been successfully applied both to formamide decomposition[33] and urea decomposition in water.[35] Briefly, we defined state A and B as the reactants ($CO_2 + H_2O$) and the product ($H_2CO_3$) basins, respectively, and formulated s and z to be s=1 in state A and s=2 in state B, as follows:

$$s(t) = \frac{e^{-\lambda D[\mathbf{R}(t),\mathbf{R}_A]} + 2\,e^{-\lambda D[\mathbf{R}(t),\mathbf{R}_B]}}{e^{-\lambda D[\mathbf{R}(t),\mathbf{R}_A]} + e^{-\lambda\,[\mathbf{R}(t),\mathbf{R}_B]}} \quad (3)$$

$$z(t) = -\frac{1}{\lambda}\log\left(\frac{e^{-\lambda D[\mathbf{R}(t),\mathbf{R}_A]} + e^{-\lambda D[\mathbf{R}(t),\mathbf{R}_B]}}{2}\right) \quad (4)$$

The key ingredient of the path CVs is the definition of distance between the atomic configuration at time $t$, $\mathbf{R}(t)$, and the reference structure, $\mathbf{R}_A$ or $\mathbf{R}_B$. Considering the different characteristics of the three media, we had to design specific distances in the three cases, which resulted in including different coordination numbers to distinguish state A and B.

In the case of the gas-phase and sc-$CO_2$ environments, we defined:

$$D[\mathbf{R}(t), \mathbf{R}_A] = \left(C_{O_wC} - C_{O_wC}^A\right)^2 \quad (5)$$

$$D[\mathbf{R}(t), \mathbf{R}_B] = \left(C_{O_wC} - C_{O_wC}^B\right)^2 \quad (6)$$

Here, the reference values for $C_{O_wC}$ is 0 in state A and 1 in state B (see Scheme 1 (a)). The coordination number, $C_{O_wC}$, was calculated as follow:

$$C_{O_wC} = \sum_{i \in C} \frac{1 - \left(\frac{r_{O_w,i}}{r_0}\right)^n}{1 - \left(\frac{r_{O_w,i}}{r_0}\right)^m} \quad (7)$$

where $r_{O_w,i}$ is the distance between the water-oxygen and the $i^{th}$ carbon, while $r_0$=2.0 Å, $m$=6, and $n$=12. In both the gas-phase and sc-$CO_2$ environments, a problem of indistinguishability of oxygens arises when $H_2CO_3$ decomposes back to $CO_2$ and $H_2O$. For this reason, we had to define the coordination number $C_{O_wC}$ based on the a priori identification of the reacting water oxygen. More specifically, the position of the water oxygen was identified as a weighted average position based on the value of the coordination number distributions, $C_{i,H}$:

$$\boldsymbol{R}_{O_w}(t) = \frac{1}{2\pi} \arctan\left[\frac{\sum_{i\in O} w_i C_{i,H} \sin(2\pi \boldsymbol{R}_i(t))}{\sum_{i\in O} w_i C_{i,H} \cos(2\pi \boldsymbol{R}_i(t))}\right] \quad (8)$$

This calculation is activated by the keyword CENTER_OF_MULTICOLVAR implemented in PLUMED 2 from version 2.3 on. Here, the $\boldsymbol{R}_i(t)$ values correspond to the position of the i$^{th}$ oxygen atom at time *t*, while $C_{i,H}$ is defined as the number of hydrogen atoms around the i$^{th}$ oxygen:

$$C_{i,H} = \sum_{j\in H} \frac{1 - \left(\frac{r_{i,j}}{r_0}\right)^n}{1 - \left(\frac{r_{i,j}}{r_0}\right)^m}, i \in O \quad (9)$$

where $r_{i,j}$ is the distance between atom *i* and atom *j*, while $r_0$, *m*, and *n* are the parameters of the switching function that were chosen equal to 1.5 Å, 6, and 12, respectively. Knowing the distribution of $C_{i,H}$, we can then assign the weights $w_i$. $\boldsymbol{R}_{O_w}$, thus, takes into account the coordinates of the oxygen atom together with those of closely coordinating hydrogens. Once the water-oxygen position is identified, we can calculate $C_{O_wC}$.

To follow the same reaction in a water environment, the distances used to build the path CVs are:

$$D[\mathbf{R}(t), \mathbf{R}_A] = (C_{CO} - C_{CO}^A)^2 \quad (10)$$

$$D[\mathbf{R}(t), \mathbf{R}_B] = (C_{CO} - C_{CO}^B)^2 \quad (11)$$

where $C_{CO}$ consists of the (rescaled) number of oxygens around the carbon atom of the only CO$_2$ present in the box:

$$C_{CO} = \sum_{i \in O} \frac{1 - \left(\frac{r_{C,i}}{r_0}\right)^n}{1 - \left(\frac{r_{C,i}}{r_0}\right)^m} - 2 \qquad (12)$$

here $r_{C,i}$ is the distance between the carbon atom and the $i^{th}$ oxygen, while $r_0$=2.0 Å, $m$=12 and $n$=24. In this fashion, the reference value for $C_{CO}$ is 0 in state A and 1 in state B (see Scheme 1 (a)).

A more natural and transparent representation of the chemistry of the reaction was then obtained by projecting the FES onto two variables CV1 and CV2. These are meant in general terms to represent the number of C-O and O-H bonds involved in the reaction, respectively.

For the gas-phase and sc-CO$_2$ (where we have only one H$_2$O), CV1 is $= C_{O_wC}$, and gives the number of carbons around the water-oxygen. To determine CV2, we calculated again the distribution of the coordination numbers of all the oxygens with all the hydrogens in the system, $C_{i,H}$, and then we counted the number of oxygens with more than 1.5 hydrogens bonded, as it follows:

$$CV2 = C_{OH} = \sum_{i \in O} 1 - \frac{1 - \left(\frac{C_{i,H}}{C_{OH}^0}\right)^N}{1 - \left(\frac{C_{i,H}}{C_{OH}^0}\right)^M} \qquad (13)$$

where N=6 and M=12 and $C_{OH}^0$=1.5. With this CV we could discriminate between state A, characterized by CV2=1, and state B, for which CV2=0.

In the case of water, we used CV1=$C_{CO}$, which is the normalized number of oxygens around the carbon atom of CO$_2$. Whereas, to describe correctly the number of hydrogens bound to any oxygen needed in the computation of CV2, we used a hydrogen-oxygen coordination distribution calculated by Voronoi tessellation as implemented by Grifoni et al.[36] Briefly, for each i-th oxygen, the hydrogen coordination number can be calculated as:

$$C_{i,H}^{water} = \sum_{j \in H} \frac{e^{-\lambda r_{i,j}}}{\sum_{m \in O} e^{-\lambda r_{m,j}}}, i \in O \qquad (14)$$

where $\lambda=4$. Similarly to the previous case, we can apply a switching function to this distribution ($C_{i,H}^{water}$) to set the OH coordination ($C_{OH}^{water}$) equal to 1 in state A and 0 in state B counting the number of oxygens with more than 1.5 hydrogens bonded, as follows:

$$CV2 = C_{OH}^{water} = \sum_{i \in O} 1 - \frac{1 - \left(\frac{C_{i,H}^{water}}{C_{OH}^0}\right)^N}{1 - \left(\frac{C_{i,H}^{water}}{C_{OH}^0}\right)^M} \qquad (15)$$

where N=6 and M=12 and $C_{OH}^0$=1.5.

**2.3. Collective Variables for the reaction $H_2CO_3$ → $HCO_3^-$ and conformers.** In acid-base reactions, such as the deprotonation of carbonic acid, one forms either a hydronium ($H_3O^+$) or a hydroxyl ($OH^-$) ion.[37-39] The structure of the solvated ions is rather elusive, as they can rapidly diffuse in the medium via a Grotthuss mechanism.[16] They are highly fluxional and the identity of the atoms involved in their structure changes continuously. As a result, it is very difficult to capture the nature of these species with an explicit analytic function of the atomic coordinates. Recently, Grifoni et al.[36] developed CVs that can follow the protonation state of the system ($s_p$) and the distance between the two conjugate acid-base sites ($s_d$). This CVs enabled us to correctly assign the reactive centers during the carbonic acid deprotonation reaction and accurately compute the corresponding reaction free energy.

Carbonic acid has three possible conformers, depending on the relative orientation of the hydroxyl groups, commonly called cis-trans (CT), trans-trans (TT), and cis-cis (CC), illustrated in Scheme 1 (b). To monitor the presence of the different conformers of $H_2CO_3$ in either the sc-$CO_2$ or water environments a more complex variable was needed than just the dihedral angles of the molecule, because of the oxygen exchange (scrambling) between $H_2CO_3$ and $CO_2$ (scCO$_2$) or $H_2O$ (aqueous). To follow this exchange, yet another collective variable, $d_{C-HC}$, had to be defined, as the distance between the carbon atom of the $H_2CO_3$ and Hc, the mid-distance of the H atoms in $H_2CO_3$, see Scheme 1 (b). Still, the identification of C and Hc is not straightforward in the sc-$CO_2$ and water environments, but rather exhibits a very dynamic behavior.

In sc-$CO_2$, the reacting $CO_2$, and hence the C center, cannot be identified *a priori*. Following the same approach as in the case of water, we defined the C center as the weighted average position based on the coordination number distribution $f_i$:

$$\boldsymbol{R_{C_{H2CO3}}}(t) = \frac{1}{2\pi} \arctan \left[ \frac{\sum_{i \in C} w_i f_i \sin(2\pi \boldsymbol{R}_i(t))}{\sum_{i \in C} w_i f_i \cos(2\pi \boldsymbol{R}_i(t))} \right] \tag{16}$$

and

$$f_i = \sum_{j \in H} \frac{1 - \left(\frac{r_{i,j}}{r_0}\right)^n}{1 - \left(\frac{r_{i,j}}{r_0}\right)^m}, i \in C \tag{17}$$

where $f_i$ represents the number of hydrogen atoms around the $i^{th}$ carbon, $r_{i,j}$ is the distance between atom $i$ and atom $j$, and $r_0$, $m$, and $n$ are the parameters of the switching function with values set to 1.7 Å, 24 and 48, respectively, such that the two basins are well separated. As a result, $\boldsymbol{R_{C_{H2CO3}}}$, represents the C center with the larger number of nearby hydrogens. Once the C center of the carbonic acid is identified, we can define as collective variable the distance between this atom and the center of the hydroxyl-hydrogens, $d_{C-Hc}$, as illustrated in Scheme 1 (b).

In water, we have an opposite scenario: whereas in this case of $CO_2$ the carbon atom is well identified since it is the only carbon present in the box, the Hc cannot be defined a priori, due to the of the rapid proton shuttling of the $H_2CO_3$ hydroxyl hydrogens. Hence, we define Hc as:

$$\boldsymbol{R_{Hc}}(t) = \frac{1}{2\pi} \arctan \left[ \frac{\sum_{i \in H} w_i g_i \sin(2\pi \boldsymbol{R_i}(t))}{\sum_{i \in H} w_i g_i \cos(2\pi \boldsymbol{R_i}(t))} \right] \quad (18)$$

and

$$g_i = \sum_{j \in C} \frac{1 - \left(\frac{r_{i,j}}{r_0}\right)^n}{1 - \left(\frac{r_{i,j}}{r_0}\right)^m}, i \in H \quad (19)$$

where $g_i$ represents the number of carbon atoms around the $i^{th}$ hydrogen and $r_{i,j}$ is the distance between atom $i$ and atom $j$, while $r_0$, $m$, and $n$ are the parameters of the switching function chosen to be 2.0 Å, 16 and 32, respectively. $\boldsymbol{R_{Hc}}$ represents the weighted average position of the hydrogens closer to the carbon atom, which is exactly the mid-distance between the two hydroxyl hydrogens (Hc). Again, once the mid-distance (H$_C$) is identified, we can define as collective variable the distance between the carbon atom and H$_C$, $d_{C-Hc}$.

Given the number of processes investigated, in Table 1 we summarize all the reported the CVs used for all the different reactions studied in this work.

**Table 1. Summary of CV used to perform each metadynamic run and to calculate the corresponding FES.**

| Reaction: | Environment | Metadynamics performed on: | FES reweighted on: |
|---|---|---|---|
| CO$_2$ + H$_2$O → H$_2$CO$_3$ | gas phase | path variables s(t) and z(t) | CV1=C$_{OwC}$ CV2=C$_{OH}$ |
| | scCO$_2$ | path variables s(t) and z(t) | CV1=C$_{Ow-C}$ CV2= C$_{OH}$ |
| | water | path variables s(t) and z(t) | CV1=C$_{CO}$ CV2=C$_{OH}^{water}$ |
| H$_2$CO$_3$ → HCO$_3$$^-$ | water | s$_P$ and s$_D$ | s$_P$ and s$_D$ |
| H$_2$CO$_3$: conformational analysis | gas phase | Φ and Ψ (see Fig. 5) | Φ and Ψ (see Fig. 5) |
| | scCO$_2$ | | CV1=C$_{OwC}$ d$_{C-Hc}$ |
| | water | | CV1=C$_{CO}$ d$_{C-Hc}$ |

**2.4. Free Energy, Internal Energy and Entropy Maps.** The internal energy and entropy maps were built following the procedure described in a recent publication by Salvalaglio and coworkers.[40] The ensemble average of the internal energy $\langle \Delta U \rangle_s$ is mapped on the CV space $s = \{CV1, CV2\}$:

$$\langle \Delta U \rangle_s = \frac{\iint U \, e^{-\beta F(s,U)} \delta(s-s') s dU}{\iint e^{-\beta F(s,U)} \delta(s-s') ds dU} \quad (20)$$

At convergence, the ensemble average of the free energy $\langle \Delta F \rangle_s$ can be written as:

$$\langle \Delta F \rangle_s = \iint e^{-\beta F(s,U)} \delta(s-s') ds dU \quad (21)$$

with,



$$e^{-\beta F} = e^{-\beta(V_{bias}+c(t))} \tag{22}$$

Here $\beta = 1/k_B T$, $V_{bias}$ is the bias deposited along the metadynamics run, and $c(t)$ is the reweighting factor that can be computed as proposed in the paper by Tiwary and Parrinello:[41]

$$c(t) = \frac{1}{\beta} \log \frac{\int e^{-\beta F(s)} ds}{\int e^{-\beta[F(s)+V_{bias}(s)]} ds} \tag{23}$$

Knowing $\langle \Delta U \rangle_s$ and $\langle \Delta F \rangle_s$ it is possible to compute the ensemble average of entropy $\langle T\Delta S \rangle_s$ as:

$$\langle T\Delta S \rangle_s = \langle \Delta U \rangle_s - \langle \Delta F \rangle_s \tag{22}$$

Quasi harmonic calculations were also carried out to compare the entropic contribution to the reaction free energy in the different environments, with the one obtained by metadynamics calculation.

## 3. RESULTS AND DISCUSSION

Figure 1 summarizes the free-energy landscapes for the formation of carbonic acid from its constituent molecules $CO_2$ and $H_2O$, in the three different environments, (I) gas phase, (II) $scCO_2$, and (III) water. These were based on the metadynamics simulations with the collective variables described in the previous section. From our unbiased ab initio molecular dynamics simulations performed for both A (reactants, $CO_2+H_2O$) and B (products, $H_2CO_3$) basins, we computed the radial distribution functions to assess the local structure in $scCO_2$ and water, see sections S1 and S2 in SI. In Fig. 2 we plot the minimum free energy path that clearly shows the barriers for the forward and reverse reactions. The gas phase reaction proceeds through a narrow, steep channel with a free-energy barrier of 51.4 kcal/mol. The gas phase

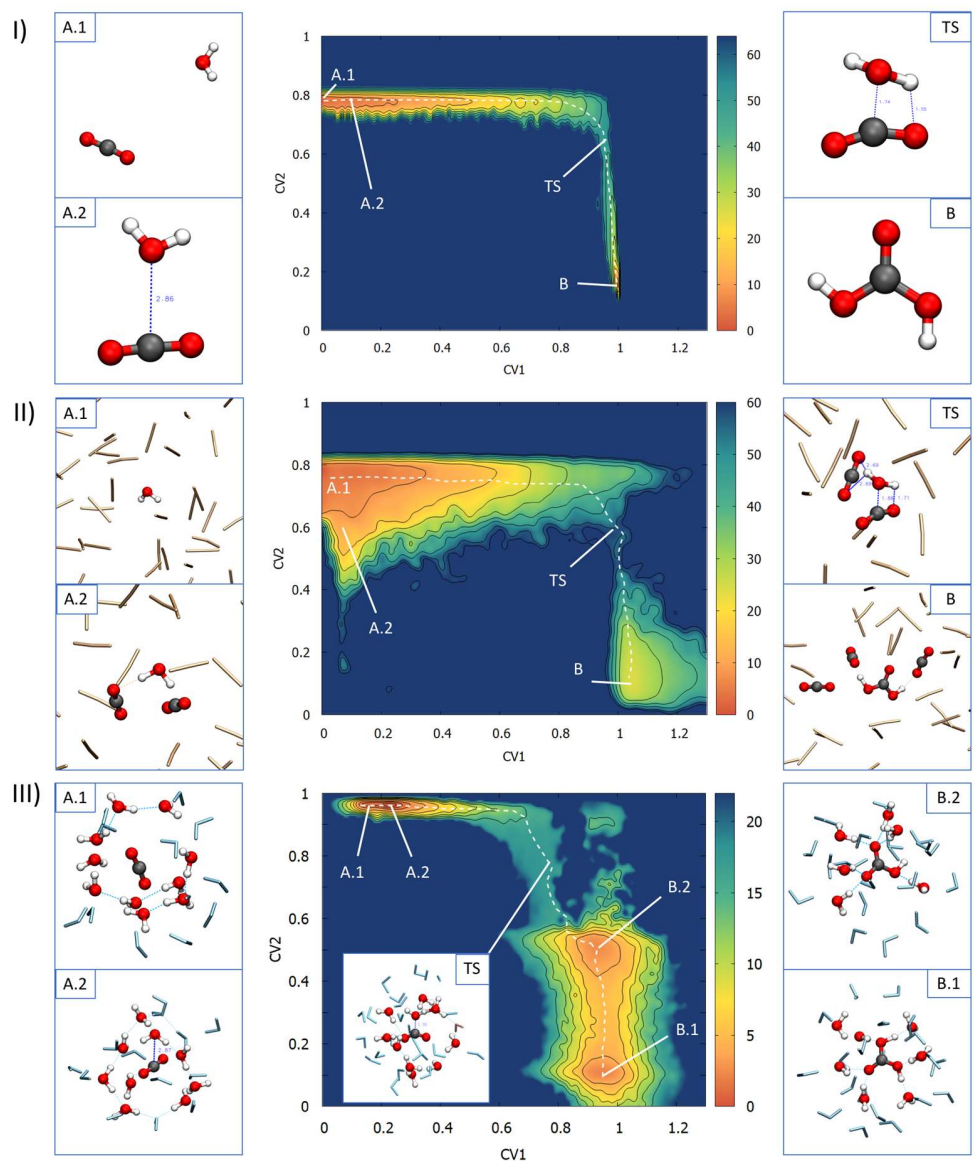

Figure 1. Free energy surface (FES) for the reaction CO + $H_2O \rightleftarrows H_2CO_3$ in (I) gas-phase, (II) sc-$CO_2$ and (III) in water. The white dashed lines depict the minimum energy path (MEP) along each FES. Units are in kcal/mol.

---

energetics of the $CO_2/H_2O$ association has been studied extensively. The geometries and relative energetics between the $CO_2\cdots H_2O$ pre-complex, transition state and product computed here are in excellent agreement with experimental measurements[42] or theoretical calculations, see for example Wight and Boldyrev,[30] Loerting et al.,[8] Kumar et al.,[10] Nguyen et al,[6] and Gallet et al.[11] and references therein. In the gas phase, the pre-complex is a loosely bound van der Waals complex (A2) with a binding energy of ~2 kcal/mol where $CO_2$ and $H_2O$ are roughly on the same plane and the C and $O_w$ centers are aligned. On the free energy surface, the pre-complex appears to be iso-energetic with the dissociated $CO_2$/$H_2O$ state. The transition state (TS) occurs after significant charge transfer from the $O_w$ to C resulting in a bent $CO_2$[43] and strong H-bonding between one of the $H_2O$ hydrogens and one of the $CO_2$ oxygens. The reverse barrier in the gas-phase of 42.3 kcal/mol is also in very good agreement with other published studies. Next, we compare how the FES in $scCO_2$ and water reactions to the gas-phase picture.

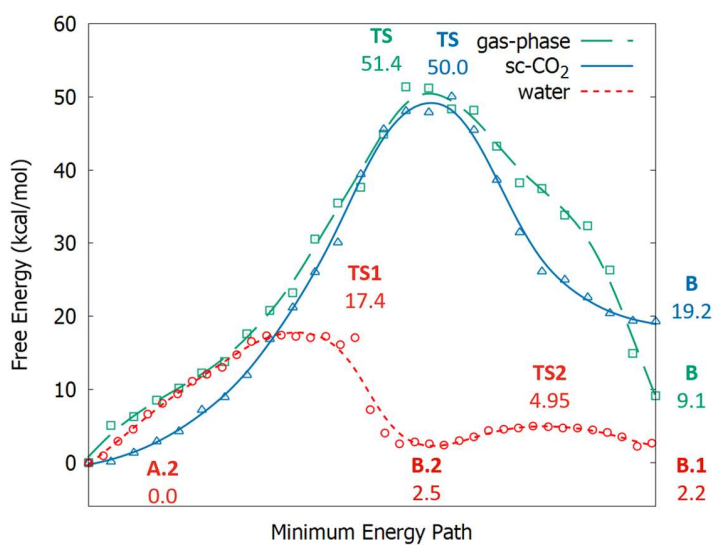

**Figure 2**. Minimum energy pathway (MEP) from the computed FES for the reaction CO + $H_2O \rightleftarrows H_2CO_3$ in (I) gas-phase, (II) sc-CO2 and (III) in water. Units are in kcal/mol.

In scCO₂, each $CO_2$ is coordinated by several others along the equatorial plane and the oxygen poles and their geometry shows instantaneous, yet significant deviations from linearity, indicative of induced dipoles attributed to the inhomogeneous near-neighbor environment.[12] The structure and dynamics of monomeric $H_2O$,[13,14] dispersed and small water clusters[14] in scCO₂ have been examined before by means of AIMD. All of the studies show that the distorted T-shaped configuration of scCO₂ is not perturbed by the presence of neither dispersed monomeric $H_2O$ nor by small $H_2O$ clusters. The $H_2O$/$CO_2$ interactions are fairly weak and the strongest interactions occur with the monomeric $H_2O$ that forms H-bonds with the oxygens of $CO_2$. Given the presence of the equatorial coordinating CO₂s at ~3.4 Å, we hypothesized that the reaction between $CO_2$ and $H_2O$ to form $H_2CO_3$ would likely have a high barrier, since $H_2O$ would have to displace the surrounding $CO_2$'s out of the equatorial positions and align with the reacting C to approach at distances < 3.0 Å. Our hypothesis was confirmed by the result of the metadynamics AIMD simulations: the computed forward free energy barrier is 50 kcal/mol, only 1.4 kcal/mol lower than the gas phase, Fig. 2. The reverse barrier for the scCO₂-mediated $H_2CO_3$ decomposition is 30.8 kcal/mol, about 10 kcal/mol lower than the gas phase.

We will now discuss the carbonic acid formation and decomposition in water. Table 2 summarizes the forward/reverse barriers for this reaction from this work, as well as a representative subset of the literature on the subject. Several studies have shown that addition of even 1 additional $H_2O$ in the gas-phase helps considerably reduce the barrier, by 18–20 kcal/mol, see Table 2. Addition of more $H_2O$ molecules (3 or 4) that micro-solvate the gas-phase system further reduce the barrier almost by half. Nonetheless, the reported results show that adding 2 or more extra $H_2O$ molecules to the systems is not enough to obtain agreement with experiments. Reasonable agreement, though, was found by Stirling and Papai[44] adopting enhanced ab initio molecular dynamics simulations on extended models. This emphasizes the importance of including the proper collective motions of the environment to correctly describe the properties of condensed phase reactions.

**Table 2. Forward/Reverse (first/second row) energy barriers of carbonic acid formation in different media and reaction free energies, activation barriers and pK$_A$ for carbonic acid dissociation in water.**

| | $CO_2 + H_2O \rightarrow H_2CO_3$ | | | | | | |
|---|---|---|---|---|---|---|---|
| | gas phase | microsolvation (n = 1,2,3 assisting H$_2$O molecules) | | | | water | sc-CO$_2$ |
| | | 1 H$_2$O | 2 H$_2$O | 3 H$_2$O | n>3 H$_2$O | | |
| **Theoretical** | | | | | | | |
| This work | 51.4 | | | | | 17.4 | 50.0 |
| | 42.3 | | | | | 14.9 | 30.8 |
| Nguyen et al.[45] | 49.2 | 15.5 | | | | | |
| | 54.0 | 5.0 | | | | | |
| Loerting et al.[46] | 52.6 | 34.7 | 31.6 | | | | |
| | 43.6 | 27.13 | 24.0 | | | | |
| Tautermann et al.[9] | 52.8 | 34.2 | | 29.1 | | | |
| | 44.8 | 27.7 | | 20.9 | | | |
| Jena and Mishra[5] | 50.6 | 31.8 | | | | | |
| | -- | -- | | | | | |
| Kumar et al.[10] | -- | 33.0 | | | | | |
| | 37.1 | | | | | | |
| Nguyen et al.[6] | 51.6 | 31.2[a] | 25.6 22.3[a] | 19.9 19.0[a] | | | |
| | 41.0 | 24.4 | 20.2 17.4[a] | -- | | | |
| | | 23.3[a] | | | | | |
| Gallet et al.[11] | 54 | 31 | 24 | | | | |
| | 38 | 19 | 18 | | | | |
| Stirling and Papai[44] | | | | | 18.8 | | |
| | | | | | 15.5 | | |
| Wang and Cao[47] | 48.9 | 33 | 21.6 | 20.3 | 19 | | |
| | 44.4 | 24.9 | 17.0 | 10.9 | 15 | | |
| **Experimental** | | | | | | | |
| Meier and Schwarzenbach[48] | | | | | | -- | |
| | | | | | | 16.1 | |
| Magid and Turbeck[49] | | | | | | 17.7 | |
| | | | | | | 14.6 | |
| Pocker and Bjorkquist[37] | | | | | | 19.5 | |
| | | | | | | -- | |
| Wang et al.[38] | | | | | | 19.3 | |
| | | | | | | 17.1 | |
| | $H_2CO_3 \rightarrow HCO_3^-$ | | | | | | |

|  | ΔG$_R$ | pK$_A$ | E$_A$ |
|---|---|---|---|
| **Theoretical** | | | |
| This work | 4.0 | 2.7 (323 K) | |
| Stirling and Papai[44] | 5.1 | 3.7 (350 K) | |
| Galib and Hanna[39,50] | 5.5 | | 9.3 |
| Wang and Cao[47] | 8.9–7.5 | | 8.6–11.1 |
| **Experimental** | | | |
| Adamczyk et al.[51] | 4.7 | 3.45 (298 K) | |
| Pines et al.[52] | 4.8 | 3.49 ( 298 K) | |

[a] Solvation correction calculated with COSMO-PCM.

The reported free energy barrier of 18.8 kcal/mol calculated at 350 K is consistent with the experimental value of 21.8 kcal/mol measured at the same temperature. The minimum energy path determined on this surface at 323 K proceeds with a barrier of 17.4 kcal/mol with the concerted addition of H$_2$O to CO$_2$ and immediate formation of an ion pair HCO$_3^-$/H$_3$O$^+$. The difference between this work and that of Stirling and Pápai is likely due to our choice of CVs, which allows for a better sampling of reactive events (several crossings/re-crossing across the transition state) and consequently a more accurate free energy surface. The same ion pair formation mechanism has been discussed before by Stirling and Pápai[44] and Adamczyk et al.[51] as the potential culprit for the unusual kinetic stability of H$_2$CO$_3$ and lower than expected *pK$_a$* values. The free energy barrier estimated for the formation of the carbonic acid is in good agreement with values measured at the same temperature, 17.4 kcal/mol vs 17.7[49] and 19.3 kcal/mol.[38]

It is worth mentioning that during the simulation we detected also the formation of the carbonate ion (CO$_3^{2-}$), whose geometry is depicted in Fig. 3. This structure corresponds to the relative minimum observed in the (1, 1) region of the CV space on the FES reported in Fig. 1.III. The CVs adopted in this metadynamics simulation were not designed to follow the equilibrium reaction between H$_2$CO$_3$, HCO$_3^-$ and CO$_3^{2-}$. Still, we obtained useful qualitative information on the relative stability of $HCO_3^-$, $H_2CO_3$, and $CO_3^{2-}$ with the latter being the least stable. A more detailed analysis of the first deprotonation reaction of $H_2CO_3$ is presented below.

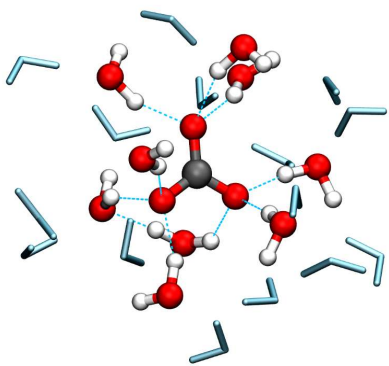

**Figure 3**. Geometry of carbonate $CO_3^{2-}$ ion in water.

To quantify the $H_2CO_3 \rightleftharpoons HCO_3^-$ equilibrium reaction, we used the CVs proposed by Grifoni et al.[36], $s_p$ and $s_d$, which have been designed specifically to follow acid-base equilibria reactions. The FES obtained is illustrated in Fig. 4 (a) together with the corresponding free energy profile along the minimum energy path, in Fig. 4 (b). During the deprotonation reaction, the system passes through an intermediate complex (B*) formed by the ionic pair $HCO_3^-/H_3O^+$ which is only 2.6 kcal/mol higher in energy than $H_2CO_3$. Only then is the pair completely separated ($s_d > 6.0$ Å) and the relative stability between carbonic acid and bicarbonate can be measured. The reaction free energy determined is equal to ~4.0 kcal/mol with a corresponding $pK_A$ of 2.7. Both values are in reasonable agreement with experiments[51,52] and previous theoretical calculations[44] (see Table 2).

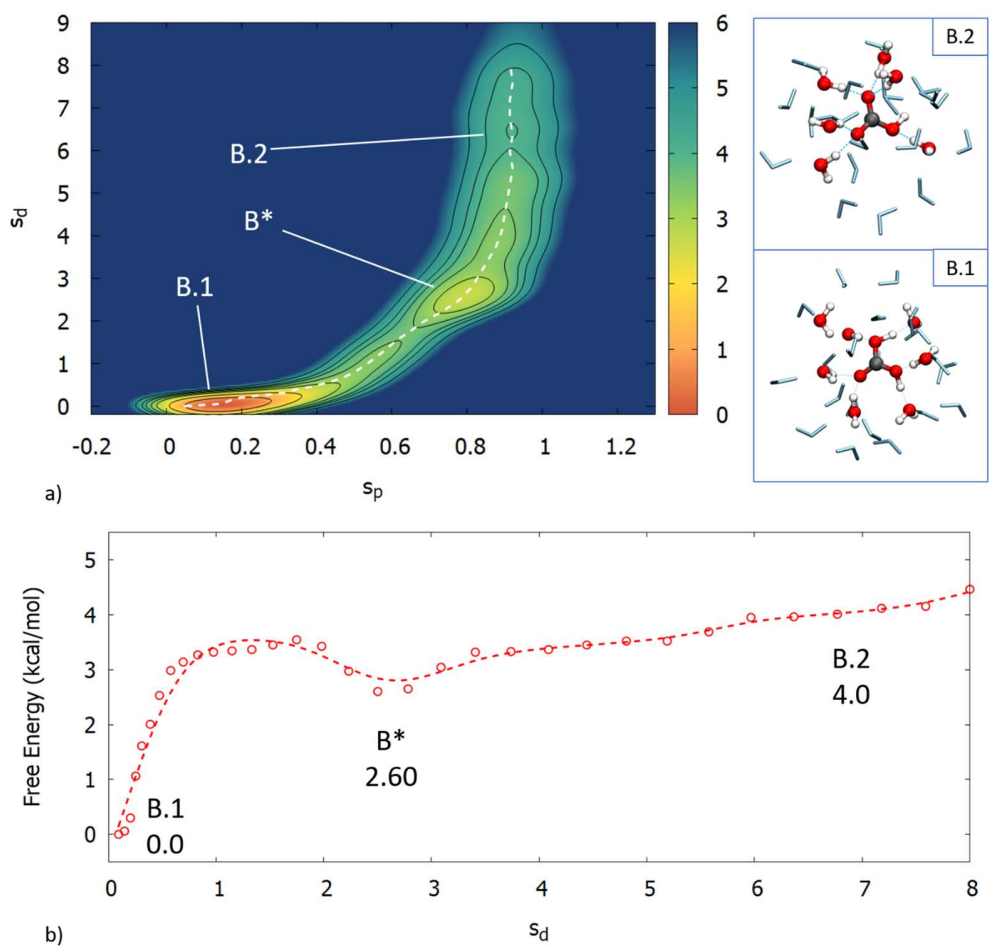

**Figure 4.** Free energy surface in water a) and MEP b) for the $HCO_3^- \leftrightarrow H_2CO_3$ reaction. B* corresponds to a metastable complex formed by the ionic couple $[HCO_3]^- \cdots H_3O^+$.

Finally, we also investigated the conformational equilibrium between the $H_2CO_3$ conformers, CC, TT, and CT (see Scheme 1 (b)) in the different environments. In gas phase, the different conformers cannot be accessed because the barriers dividing them are much larger than 1 $k_BT$ (0.642 kcal/mol at 323 K). Spontaneous conformational isomerization is thus unlikely, and no conformational changes were detected during the metadynamics simulation of carbonic acid formation in the gas phase. To further interrogate this conformational transformation, we carried out a second metadynamics run, placing a bias along the dihedral angles Φ and Ψ. With this approach, we recovered the free energy surface that connects the three states CC, CT, and TT (see Fig. 5). CC is the global minimum, with CT being ~1.5 kcal/mol higher, whereas the TT metastable state is found at ~8.5 kcal/mol in our FES. The TS dividing CC and CT is ~9 kcal/mol. Loerting and Bernard report similar values with CT 1–2 kcal/mol and TT is ~10 kcal/mol higher than CC, while the TS can be found at ~12 kcal/mol. More recently, also Schwedtfeger and Mazziotti[29] reported that CT is 1.2 kcal/mol higher in energy than CC and that the barrier dividing CC and CT is of 9.5 kcal/mol. Bernard et al.[31] carried out spectroscopy measurements that confirmed this theoretical results. In particular, they have been able to detect both conformers (CC and CT) in the gas phase at 210 K with a 1:10 ratio between CC and CT which corresponds to a difference in Gibbs energies of 1 kcal/mol.

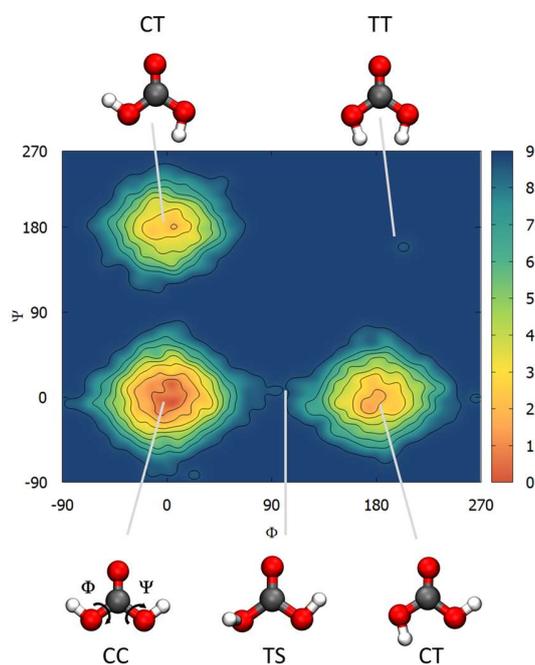

**Figure 5**. Free energy surface of the CC ↔ CT ↔ TT isomerism of $H_2CO_3$ in gas phase.

In Fig. 6 we report the results of the conformational isomerism analysis in scCO$_2$ and in H$_2$O. In both environments, we found that the barrier dividing conformers CC and CT is much lower than the barrier computed in gas phase and comparable to 1 k$_B$T. In scCO$_2$, we found CC and CT to be almost equi-energetic, while in H$_2$O, the CC structure is ~4.0 kcal/mol higher. This is in contrast with gas phase results predicting the CC isomer to be the global minimum. On the contrary, we found the CT conformer to be more stable than CC by ~2.4 kcal/mol, and this agrees well with the experimental observations of Loerting and Bernard[7] where only the CT conformer has been detected in water vapor environment.

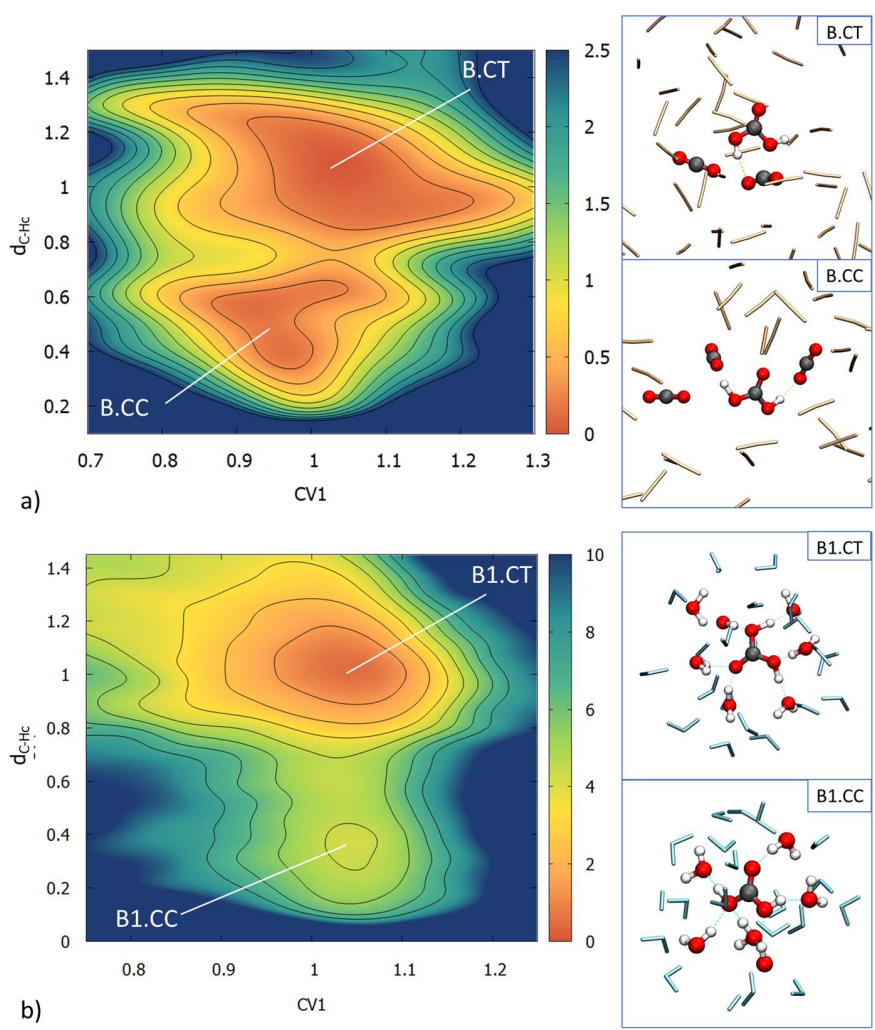

**Figure 6.** Free energy surface of the CT ↔ TT isomerism of H$_2$CO$_3$ in (a) sc-CO$_2$ and (b) water.

Finally, the entropy contributions (<T∆S>) and internal energy (<∆U>) to the free energy (<∆F>) maps are reported in Fig. 7 for the three different media. The reaction free energy, energy and entropy calculated using eqs. (20)–(24) are reported in Table 3. Results from quasi-harmonic approximation analysis[14,53,54] were carried out for comparison and both methods provide very similar estimates with errors < 20%. In gas and sc-$CO_2$ phase, <∆U> increases as the systems depart from the minima. Contrarily, in water, this behavior is true only for $H_2CO_3$ but not for $CO_2$ + $H_2O$, where the internal energy and entropy terms cancel out. Further analysis of the entropic contributions reveals a noticeable difference in the entropy between the reactants and the products, because the formation of $H_2CO_3$ in all media reduces the degrees of freedom and hence the entropy term. This difference is significantly larger in the condensed phases compared to the gas-phase. However, while in water this effect is balanced out by the enthalpic stabilization of $H_2CO_3$ through the network of water hydrogen bonds, in sc-$CO_2$ there is no stabilization of the product and hence $H_2CO_3$ formation is less probable.

**Table 3. Free energy $\langle \Delta F \rangle$, internal energy $\langle \Delta U \rangle$ and entropy $\langle T \Delta S \rangle$ between state A and state B (B.1 in water) in kcal/mol adopting quasi-harmonic approximation and metadynamics.**

| gas-phase | | | sc-$CO_2$ | | | water | | |
|---|---|---|---|---|---|---|---|---|
| \multicolumn{9}{c}{Quasi-harmonic approximation} | | | | | | | | |
| $\langle \Delta F \rangle$ | $\langle \Delta U \rangle$ | $\langle T \Delta S \rangle$ | $\langle \Delta F \rangle$ | $\langle \Delta U \rangle$ | $\langle T \Delta S \rangle$ | $\langle \Delta F \rangle$ | $\langle \Delta U \rangle$ | $\langle T \Delta S \rangle$ |
| 11.7 | 7.0 | -4.7 | 19.0 | 1.3 | -17.7 | 0.2 | -13.6 | -13.8 |
| \multicolumn{9}{c}{Metadynamics} | | | | | | | | |
| $\langle \Delta F \rangle$ | $\langle \Delta U \rangle$ | $\langle T \Delta S \rangle$ | $\langle \Delta F \rangle$ | $\langle \Delta U \rangle$ | $\langle T \Delta S \rangle$ | $\langle \Delta F \rangle$ | $\langle \Delta U \rangle$ | $\langle T \Delta S \rangle$ |
| 9.8 | 8.9 | -1.7 | 23.2 | 1.6 | -22.0 | 0.2 | -13.5 | -13.6 |

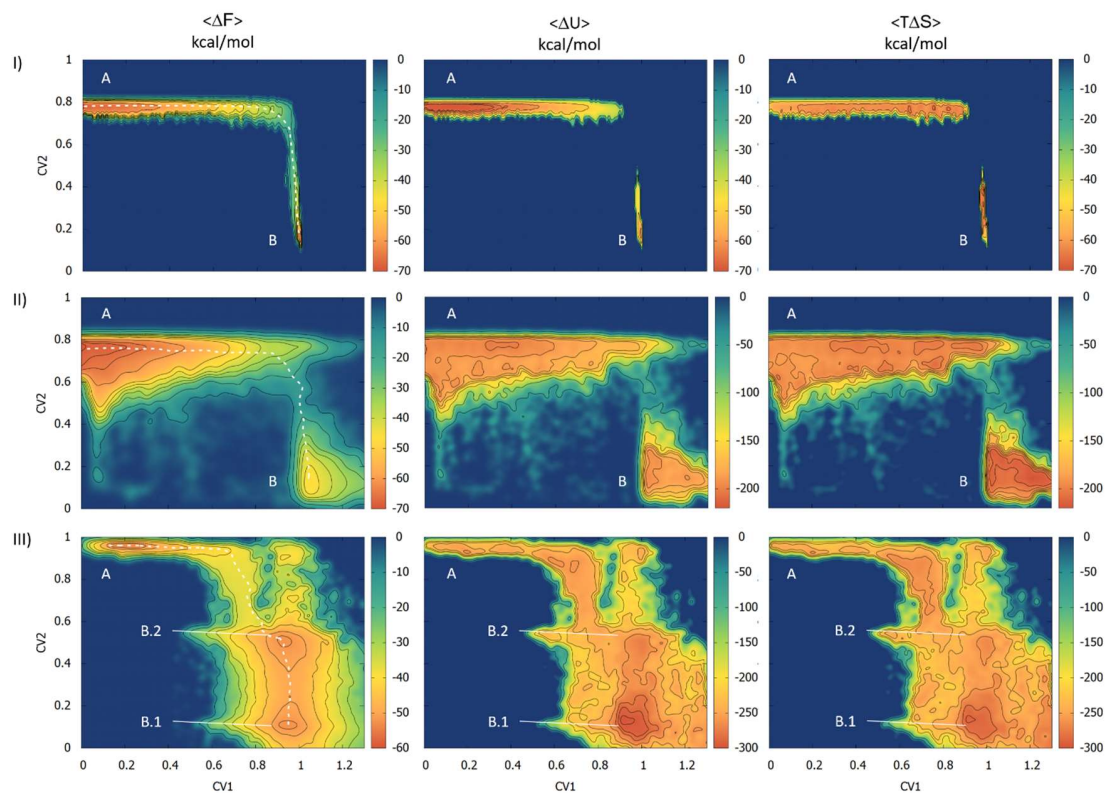

**Figure 7.** Free energy $\langle \Delta F \rangle$, internal energy $\langle \Delta U \rangle$ and entropy $\langle T\Delta S \rangle$ maps as function of the collective variables CV1 and CV2 for the reaction $CO_2 + H_2O \rightarrow H_2CO_3$ in: (I) gas-phase, (II) sc-$CO_2$ and (III) water (III).

Following the free energy decomposition maps in Fig. 7, it becomes obvious that the role of entropy in the transition state in the three different environments is very distinct. In the gas phase, the entropy term is essentially constant along the reaction path up to the transition point and it only involves the relative rotation of the reactants about the C-$O_w$ distance. As a result, the activation energy $\Delta U^\dagger$ is very similar to the activation free energy $\Delta F^\dagger$ (see Fig. 8). In contrast, in scCO$_2$, while the activation free energy is similar to the gas phase, the activation energy is appreciably larger ~ 150 kcal/mol, and compensated by an increased activation entropy, owing to the disruption of local $CO_2$ structure by $H_2O$ at the TS. In water, the overall picture is completely different: the microsolvation helps lowering the overall activation free energy but also stabilize the TS as the charge separation ensues, by providing multiple channels for proton shuttling. The same trend holds true for the product basin, where the proton shuttling and formation of hydronium species ($H_3O^+$) helps stabilize the formation of the conjugate bases ($HCO_3^-$ and $CO_3^{2-}$), but also strongly contributing to a large entropy term at both the TS and the products.

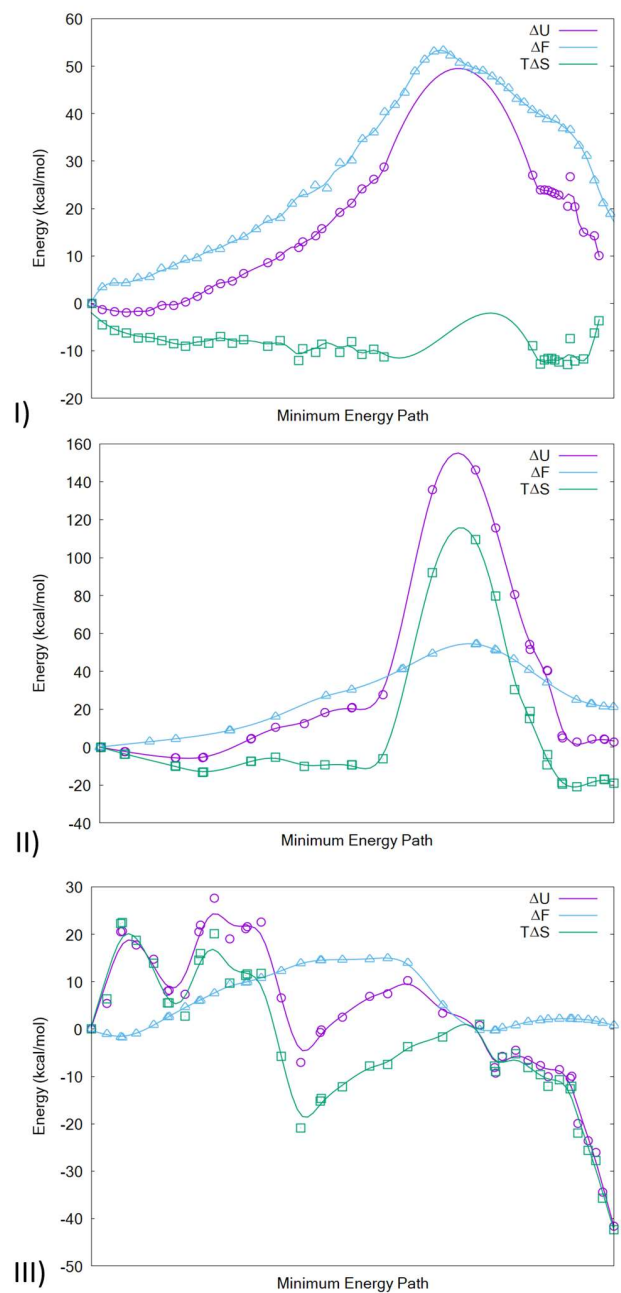

**Figure 8.** Free energy $\langle \Delta F \rangle$, internal energy $\langle \Delta U \rangle$ and entropy $\langle T\Delta S \rangle$ profiles projected on the MEP for the reaction $CO_2 + H_2O \rightarrow H_2CO_3$ in: (I) gas-phase, (II) sc-$CO_2$ and (III) water (III).

## 4. OUTLOOK

In this work, we probed the collective solvent motions that impact the reaction profile of a prototypical and environmentally relevant system, $H_2O + CO_2$. $CO_2$ chemistry and reactivity constitutes an active research direction in many fields, such as separations, catalysis and carbon remediation. The $CO_2/H_2O$ interaction is invariably and readily assumed to lead to carbonic acid formation and subsequent decomposition to its carbonate anions regardless of the nature of the solvent, i.e., water or $scCO_2$. The low solubility of water in $scCO_2$ and solvent microstructure create a still unexplored reactive environment that is very different from the aqueous phase.[4] Our simulations present evidence that the prototypical reaction in $scCO_2$ has an activation barrier as high as that in gas phase, a notion that was never before entertained in the literature. However, decomposition of the free energy landscape shows that while the enthalpy and entropy terms are comparable in $scCO_2$, they are markedly different than in gas phase. Formation of $H_2CO_3$ in $scCO_2$ is thermodynamically unfavorable, but once formed, below water saturation, it will most likely be preserved in its undissociated form. In water, the activation barrier is considerably lower, and entropy stabilizes both the transition state and the products by providing multiple channels for $H^+$ shuttling and exchange. Based on these results, we posit that the product of $H_2O$ and $CO_2$ and speciation in $scCO_2$ (bulk solvent) is likely to be very different than the same reaction in water. However, at a solid/liquid interface, $CO_2$ activation can also happen without a direct interaction with $H_2O$ and still result in formation of carbonate species.[55]

In closing, this work highlights the importance of sampling methods[56] which capture the conformational and configurational complexity of the solvent media as well as the potential side reactions related to reactants and products. In this case, well-tailored collective variables are able to represent the overall reactive landscape and reveal novel insights into solvent effects and unanticipated reactivity.

## ASSOCIATED CONTENT

**Supporting Information**. Supporting Information is available free of charge. The file contains the following information:

1. Radial Distribution Functions: 1.1 Supercritical $CO_2$ (sc$CO_2$); 1.2 Water

2. Metadynamics Convergence.

## AUTHOR INFORMATION

**Corresponding Author**

Vassiliki-Alexandra Glezakou, Vanda.Glezakou@pnnl.gov

**Author Contributions**

D.P. designed the collective variables and performed the majority of metadynamics simulations and analyses and E.G. provided technical assistance with the Grifoni variables. V.-A.G. scoped the research and together with R.R. performed some of the simulations. All authors contributed to the writing of the manuscript and have given approval to the final version of the manuscript.

## ACKNOWLEDGMENTS

V.-A.G. and R.R. were supported by the U.S. Department of Energy (DOE), Office of Science, Office of Basic Energy Sciences, Division of Chemistry, Geochemistry and Biological Sciences, and located at Pacific Northwest National Laboratory (PNNL). Computational resources were provided by National Energy Research Scientific Computing Center (NERSC), a DOE Office of Science User Facility located at Lawrence Berkeley National Laboratory (LBNL). PNNL is operated by Battelle for the US Department of Energy under Contract DE-AC05-76RL01830. D.P., E.G. and M.P. thankfully acknowledge the financial support provided by the NCCR MARVEL, funded by the Swiss National Science Foundation, and the European Union Grant No. ERC-2014-AdG-670227/VARMET. Computational resources were provided by the Swiss National Supercomputing Centre (CSCS) under project IDs p503, s768 and s910.

**ABBREVIATIONS**